\documentclass[pre,twocolumn,aps,showpacs,amsmath,amssymb]{revtex4-1}

\usepackage{color}
\usepackage{bm}
\usepackage{graphicx}
\usepackage{braket}

\usepackage[plainpages=false,pdfpagelabels,colorlinks=true,linkcolor=red,urlcolor=blue,citecolor=blue,pdftitle={Title},pdfauthor={},pdfdisplaydoctitle=true,pdfduplex=DuplexFlipLongEdge]{hyperref}

\def\Tr{\mbox{Tr}\,}

\newcommand\ba{\begin{eqnarray}}
\newcommand\ea{\end{eqnarray}}
\newcommand\be{\begin{equation}}
\newcommand\ee{\end{equation}}

\begin{document}
\title{Eigenstate entanglement entropy in $PT$ invariant non-Hermitian system}
\author{Ranjan Modak$^1$ and Bhabani Prasad Mandal$^1$} 
\affiliation{$^1$ Department of Physics, Banaras Hindu University, Varanasi-221005, India}

\begin{abstract}
Much has been learned    about universal properties of the eigenstate entanglement entropy for 
one-dimensional lattice models, which is described by a Hermitian Hamiltonian. 
While very less of it has been understood for non-Hermitian systems. In the present work we study a 
 non-Hermitian,  non-interacting model of fermions which is invariant under combined $PT$ transformation. 
Our models show a phase transition from $PT$ unbroken phase to broken phase as we tune the hermiticity breaking parameter. 
Entanglement entropy of such systems can be defined in two different ways, depending on whether we 
consider only right (or equivalently only left) eigenstates or a combination of both left and right eigenstates which form a complete set of bi-orthonormal eigenstates. 
We demonstrate  that the entanglement entropy of the ground state and also of the typical excited states 
show some unique features in both of these phases of the system. 
Most strikingly, entanglement entropy obtained taking a combination of both left and right eigenstates shows an exponential
divergence with system size  at the transition point. While in the $PT$-unbroken phase, the entanglement entropy obtained from only the right 
(or equivalently left) eigenstates shows identical behavior as of an equivalent Hermitian system which is connected to the non-Hermitian system by a similarity transformation.

\end{abstract}
\maketitle

\section{Introduction}

Entanglement is a property of the quantum systems, providing unique ways of characterizing
quantum many-body systems~\cite{eisert.2010,vlatko.2008}.  The correlations between two entangled quantum systems that are in
an overall pure state cannot be explained by a local classical
theory~\cite{rosen.1935}.  Studies of entanglement indicators have given  insights into
properties of ground states ~\cite{vidal.2003,eisert.2002}, quantum phase transitions ~\cite{michael.2002,osterloh2002scaling}, and highly excited
 eigenstates that exhibit eigenstate thermalization ~\cite{deutsch.1991,rigol.2008,srednicki1994,d2016quantum}. 
Different measures of entanglement also have been brought lots of attention 
in the context of black hole physics ~\cite{srednicki.1993,rafael.1986}, holography~\cite{tadashi.2006,lewkowycz2013generalized}, and quantum information  scrambling in non-equilibrium
quantum dynamics ~\cite{lewis2019unifying,alba2019quantum,Modak_2020}. Recently, thanks to the  advancements of ultracold atoms in an optical lattice, 
the  measurements of an entanglement has been realized even in experiments ~\cite{islam2015measuring,kaufman2016quantum}.

On the other hand, in recent days the study of non-Hermitian systems, such as open systems or dissipative systems with gain and loss ~\cite{xu2016topological,carmichael.1993,ashvin.2019,dimitry.2016,masahito.2017}, has revealed  various intriguing  phenomena that do not exist in Hermitian systems. For an example, the complex energy
spectra of non-Hermitian systems are theoretically predicted
to host  bulk Fermi arcs ~\cite{moors.2019,emil.2018,emil.2019,yang.2019,PhysRevB.99.121101},
which has been also realized in experiments ~\cite{chan.2018}. Also,  
there is a growing interest  to extend the idea of topological Bloch theory developed in Hermitian systems to
non-Hermitian Hamiltonian~\cite{zhong.2019,zhong.2018,yositake.2019,
PhysRevLett.123.016805,lee2018tidal}.

Among a large class of non-Hermitian system, if a system  is invariant under combined parity and time-reversal ($PT$ ) 
operations, they can have purely real spectra for a  finite range of  parameters~\cite{bender2002complex,bender1998real}.
The hermiticity property which is sufficient to ensure the real spectrum of the Hamiltonian in usual quantum mechanics  is replaced by $PT$ symmetry in the case of non-Hermitian systems. 
Although the spectrum of such systems may be completely  real, the eigenstates may not form orthonormal set and may not have positive definite norms. Because of these the probabilistic interpretation of quantum theories fails and   the time-evolution
of the corresponding quantum systems becomes non-unitary. 
Later a consistent quantum theory with  complete real spectrum, unitary time evolution and probabilistic interpretation for $PT$ symmetric non-Hermitian systems has been developed in a modified Hilbert space equipped with a positive definite $CPT$ inner product\footnote { Consistent quantum theories with a bigger class of non-Hermitian systems described by pseudo-Hermitian Hamiltonians have also been formulated in the similar fashion\cite{ali,bender2007making}}.  $C$ is an additional symmetry associated with every $PT$   symmetric non-Hermitian systems. Because of this exciting realization  the research in non-Hermitian systems have received a huge boost over the past two decades \cite{bender2007making}.  $PT$ symmetric non-Hermitian systems have found numerous applications in various branches of physics and interdisciplinary ares \cite{app1,app2,app3,app4,app5,app6,app7,app8,app9,app10,app11,app12} and some of the predictions of non-Hermitian theories are experimentally observed \cite{expobs1,ruter2010observation,el2007theory,expobs4}. Another important aspects
is that  such PT symmetric non-Hermitian systems  generally exhibit a phase transition ( or more appropriately  a $PT$ breaking transition ) that separates two parametric regions (i) region of the unbroken $PT$ symmetry in which the entire spectrum is real and eigenstates of the system respects $PT$ symmetry and (ii) a region of broken $PT$ symmetry in which the whole spectrum (or a part of it) appears  as complex conjugate pairs and eigenstates of the Hamiltonian do not respect $PT$ symmetry\cite{ptphase1,ptphase2,ptphase3,ptphase4,ptphase5,ptphase6,ptphase7}. 

One of the most popular example of $PT$ symmetric systems are  open systems with balanced gain and loss  ~\cite{lee2020many,khare2000pt,mandal2013pt,raval2019deconfinement,liu2020exact}.
 Typically, in such a system, the parity  denotes a reflection symmetry
in its spatial arrangement, and when balanced gain and loss  leads to non-Hermiticity. 
Usually for small gain and loss rates, the eigenvalues of the
$PT$ -symmetric Hamiltonian describing such a system remain real, however, when the strength of the gain (or
loss) exceeds a value known as the $PT$ -symmetry breaking threshold, two or more of its eigenvalues become
degenerate and then complex-conjugate pairs. This emergence of complex conjugate eigenvalues is a signature of  $PT$
symmetry breaking.   Recent developments in the fabrication techniques of optical devices can allow one to  create and control
arrays of coupled optical waveguides, and the couplings of these arrays can be tuned to match the dynamics
of a large variety of different tight-binding Hamiltonian~\cite{peschel1998optical,christodoulides2003discretizing} and also  controlled loss and gain can  be implemented
relatively straightforwardly, allowing the observable dynamics to extend into the non-Hermitian realm~\cite{ruter2010observation,el2007theory}.

In this work, our main aim is to understand entanglement properties of the eigenstate of non-Hermitian $PT$ symmetric systems. 
In general quantum correlation is an extremely useful tool to detect different phases as well as the phase transition.  
Specially, for 1D systems on lattice, the scaling of entanglement entropy with system size gives lots of insights about the system. In order  to distinguish between gapless and gapped phases of a system ~\cite{PhysRevA.90.032301} or detect localization-delocalization transition
~\cite{RevModPhys.91.021001}, 
the eigenstate entanglement entropy is one of the most popular diagnostic. 
Here, we show that the  entanglement entropy also can be used as a probe to detect different phases of $PT$ invariant system as well as $PT$ transition. Since, the Non-Hermitian systems have two types of eigenvectors (left and right), we define the entanglement entropy 
in two different ways ~\cite{herviou2019entanglement}, depending on whether we consider only right (or equivalently only left) eigenstates or a combination of both left and right eigenstates. We find that the  entanglement entropy obtained taking a combination of both  left and right eigenstates diverges exponentially with system size  at the transition point. While in the PT-unbroken phase, the entanglement entropy obtained from only the right (or equivalently left) eigenstate shows identical  behavior as of a Hermitian system.

The paper is organized as follows: In Sec.~\ref{secII}, we introduce the non-Hermitian lattice model which is invariant under 
$PT$ transformation. Next we discuss our analytical understanding for the $2\times 2$ model in Sec.~\ref{secIII}. In Sec.~\ref{secIV}
we numerically investigate the $PT$ transition point and Sec.~\ref{secV} is devoted for the analysis of entanglement entropy 
of the ground state as well as a typical excited state. Finally, in Sec.~\ref{secVI} we summarize our results.

\section{Model}\label{secII}
We study non-interacting fermions in 1D lattice   with open boundary. The system is 
described by the following   Hamiltonian, 
\begin{eqnarray}
 {H_0}=-\sum _{j=1}^{L-1}(\hat{c}^{\dag}_j\hat{c}^{}_{j+1}+\text{H.c.}) \nonumber \\
 \label{hamiltonian}
\end{eqnarray}
where $\hat{c}^{\dag}_j$ ($\hat{c}_{j}$) is the fermionic creation (annihilation) operator at site $j$, 
which satisfies standard anti-commutation relations. $L$ is the size of the system, which we set to be an even 
number for all our calculations ( we choose the lattice spacing as unity). 

In order to make the Hamiltonian  $PT$ symmetric and non-Hermitian, we add a local  term at site $L/2$ and $L/2+1$. 
The $PT$ symmetric Hamiltonian reads as, 

\begin{eqnarray}
H=H_0+i\gamma (\hat{n}_{L/2}-\hat{n}_{L/2+1}) \nonumber \\
\label{hamiltonian_pt}
\end{eqnarray}

where, $\hat{n}_j=\hat{c}^{\dag}_j\hat{c}_{j}$ is the number operator and $\gamma$ is identified as the hermiticity breaking 
parameter. While under Parity transformation $c_j \to c_{L-j+1}$, Time reversal symmetry operation changes $i \to -i$. 
Hence, $H$ remains invariant under $PT$ transformation, which implies $[H,PT]=0$. 

 For non-zero values of $\gamma$, $H$ is non-Hermitian. Hence, its left eigenvectors $|L_n\rangle$ and $R_n\rangle$ are not the same. 
However, $H$ is diagonalizable, and $H=\sum_{n}\epsilon_n |R_n\rangle\langle L_n|$ \textcolor{black}{ with $\langle L_n|R_m\rangle =\delta _{mn}$ and $\langle R_n|R_n\rangle =1$.} $\epsilon_n$ can be identified as 
single-particle energy eigenvalues of the system, which in general is  complex. If the Hamiltonian $H$ goes through a $PT$ phase transition, then 
in the $PT$ symmetric phase, the $\epsilon_n$ s
remain real. On the other hand, in the broken $PT$ phase $\epsilon_n$ s are complex. 
\textcolor{black}{
We also verified our results for another model where we have added the hermiticity breaking terms in a $PT$ invariant way in four sites, i.e. described 
by following Hamiltonian
\begin{eqnarray}
\tilde{H}=H_0+i\gamma \sum_{j=L/2-(r-1)}^{L/2+r}(-1)^j\hat{n}_j ,\nonumber \\
\label{hamiltonian_pt4}
\end{eqnarray}
where we chose $r=2$. We also like to point out that in the absence of hermiticity breaking term, our model is same as nearest-neighbor fermionic ``tight-binding'' model, 
which is possibly the simplest exactly solvable model one can write down in the context of condensed 
matter physics. Also, we believe 
that  a variant our Hamiltonian with gain
and loss is experimentally realizable in an ultracold
fermionic system ~\cite{murakami.2019}.}
Note that for all  the many-body calculations, we chose to work at half-filling.

\section{Analytical results: $2 \times 2$ matrix example}\label{secIII}
In this section we restrict the  Hamiltonian $H$   to only a lattice of  two sites. Our aim is to analytically solve the $2 \times 2$ 
matrix to gather some insights about this model.  The Hamiltonian $H$ ~\eqref{hamiltonian_pt} is represented in the matrix form as 
\begin{eqnarray}
H^{2\times 2}=
 \begin{pmatrix}
 i\gamma & -1 \\
 -1 & -i\gamma 
\end{pmatrix}
\label{h2cross2}
\end{eqnarray}

There exists various representation of the parity operator and 
we define the parity operator for this model as,
\begin{eqnarray}
P=
 \begin{pmatrix}
 0 & 1 \\
 1 & 0 
\end{pmatrix}
\end{eqnarray}
since $P$ transforms  $\begin{pmatrix}
 1 \\
 0
\end{pmatrix}$
to $\begin{pmatrix}
 0 \\
 1
\end{pmatrix}$ 
and vice versa. It is an optimal choice for our case.  It is straightforward to check that the Hamiltonian in Eq. \ref{h2cross2} is $PT$ invariant.
In the first step, we evaluate the eigenvalues of this $2\times 2$ matrix, which is, 
\begin{eqnarray}
 E_{\pm}=\pm\sqrt{(1-\gamma ^2)}
\end{eqnarray}
It implies that for $\gamma > 1$ ($\gamma < 1$),  $E_{\pm}$ is completely complex (real). 
\textcolor{black}{Also, it is starightforward to check that for $\gamma <1$,
$|E_{+}\rangle=\frac{1}{\sqrt{2\cos\alpha}}
 \begin{pmatrix}
 e^{i\alpha/2} \\
 e^{-i\alpha/2}
\end{pmatrix}$
and $|E_{-}\rangle=\frac{i}{\sqrt{2\cos\alpha}}
 \begin{pmatrix}
 e^{-i\alpha/2} \\
 -e^{i\alpha/2}
\end{pmatrix}$
are simultaneous eigenstates of $H^{2\times 2}$ ~\cite{bender2004} and $PT$, where
$\sin\alpha=-\gamma$.
Hence, one can conclude that  the $PT$ transition occurs  for  $H^{2
\times 2}$ matrix model at $\gamma=1$, and
$PT$ symmetric phase corresponds to $\gamma<1$.}. In the $PT$ broken phase, for $\gamma >1$
$\alpha $ becomes complex and hence the eigenstates $|E_+\rangle$ and $|E_-\rangle$ are not eigenstate of $PT$.

Now we go one step ahead, and 
construct a new linear operator $C$ that commutes with both $H^{2\times 2}$ and $PT$ . The operator  $C$ 
for $H^{2 \times 2}$ matrix   turns out to be ~\cite{bender2005introduction,bender2007making}, 
\begin{eqnarray}
C=\frac{1}{\sqrt{1-\gamma^{2}}}
 \begin{pmatrix}
 -i\gamma & 1 \\
 1 & i\gamma
\end{pmatrix}
\end{eqnarray}
A more general way to represent the $C$
operator is to express it generically $C=e^{Q}P$. 
It has been shown that the square root of the positive operator $e^{Q}$ can be used
to construct a similarity transformation that maps a non-Hermitian $PT$ -symmetric
Hamiltonian $H$ to an equivalent Hermitian Hamiltonian $h$ ~\cite{bender2001quantum}, where 
\begin{eqnarray}
 h=e^{-Q/2}He^{Q/2}
\end{eqnarray}
For the $H^{2\times 2}$ matrix model, the equivalent Hermitian matrix $h^{2\times 2}$ will be, 
\begin{eqnarray}
h^{2\times 2}=
 \begin{pmatrix}
 0 & -\sqrt{1-\gamma^{2}} \\
 -\sqrt{1-\gamma^{2}} & 0
\end{pmatrix}
\label{h2times2}
\end{eqnarray}

Note that $h^{2\times 2}$ is equivalent to $H^{2 \times 2}$ because it has the same eigenvalues as $H^{2 \times 2}$. However, the eigenvectors of $H^{2 \times 2}$ and $h^{2\times 2}$ are not the same, they are related to each other with a similarity transformation \textcolor{black}{(see Appendix~\ref{appendixII} for the detail derivation 
of the equivalent Hermitian matrix $h^{4\times 4}$ for the $4\times 4$ non-Hermitian Hamiltonian)}.

\section{PT transition}\label{secIV}

\begin{figure}
\includegraphics[width=0.46\textwidth]{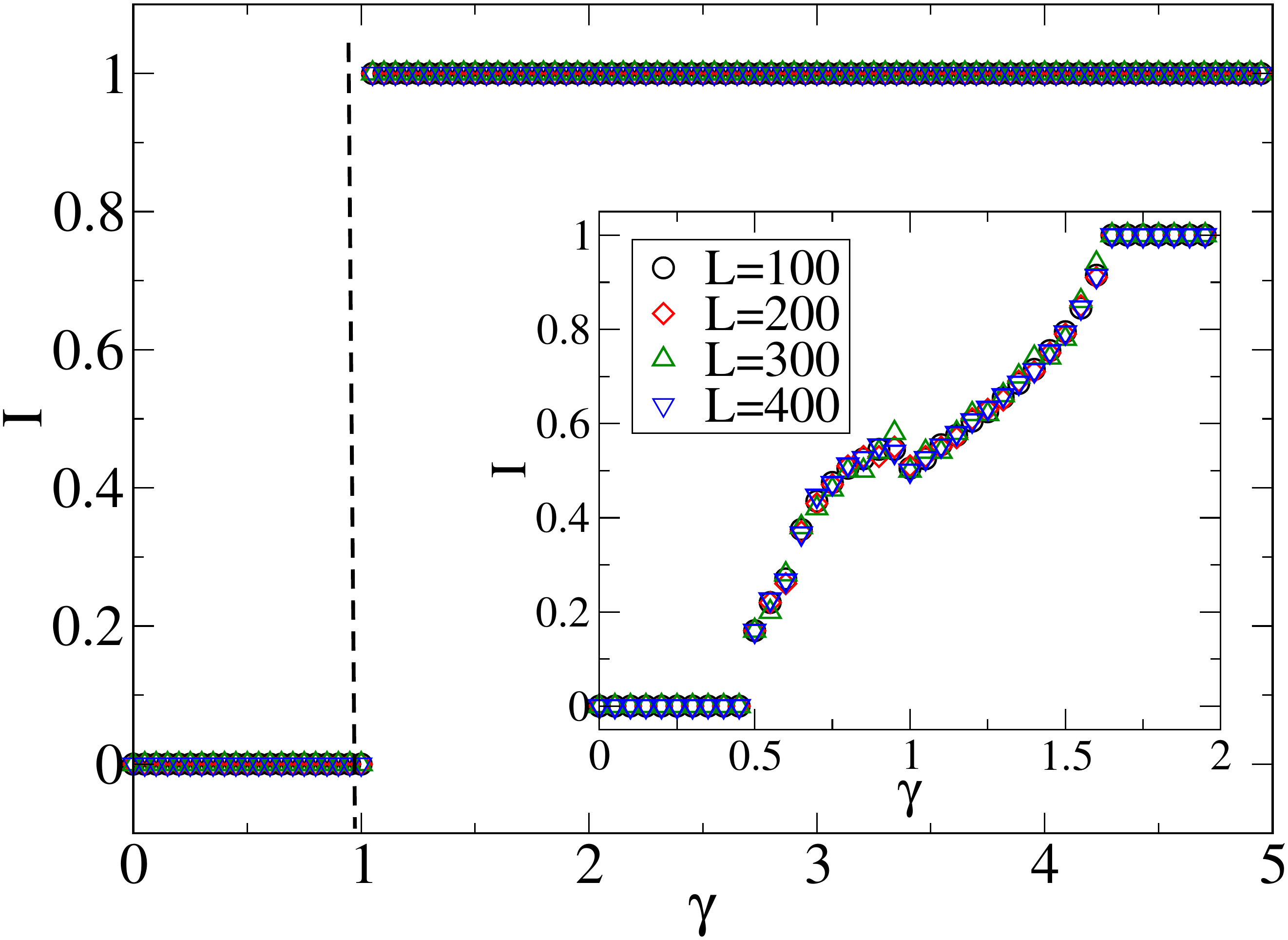}
\caption{Variation of the fraction of single-particle complex eigenvalues $I$
as a function of the hermiticity breaking parameter
$\gamma$ for $L=100$, 200, 300, and 400 for the Hamiltonian $H$
~\eqref{hamiltonian_pt}. Inset shows the variation of the fraction of
single-particle complex
eigenvalues $I$ as a function $\gamma$, where the hermiticity breaking
terms has been added in 4 sites [see Hamiltonian $\tilde{H}$ ~\eqref{hamiltonian_pt4}. }
\label{fig1}
\end{figure}

In this section, we investigate the fate of $PT$ transition for the Hamiltonian ~\eqref{hamiltonian_pt} and ~\eqref{hamiltonian_pt4} in the thermodynamic limit.   In the previous section,  we had showed analytically for the 2 sites version of  the Hamiltonian  ~\eqref{hamiltonian_pt}, the $PT$ transition occurs  at $\gamma=1$. Here we numerically diagonalize the Hamiltonian $H$ and $\tilde{H}$ for different values of $L$ to obtain all the energy eigenvalues. 
 In order to characterize $PT$ transition, we plot the fraction of complex eigenvalues $I$ as a function of $\gamma$.
We expect that in the $PT$ symmetric (unbroken) phase that ratio should be zero, whereas in the $PT$ broken phase the value of $I$ should be non-zero. 
 Figure.~\ref{fig1} (main panel) shows the variation of $I$ as a function of $\gamma$ for different values of $L$ for the Hamiltonian $H$. We see 
$I$ jumps from $0$ to $1$ ( it implies that all eigenvalues become complex in the $PT$ broken phase) at  $\gamma=1$, which concludes a clear signature $PT$ phase transition in this model, where
$PT$ broken (unbroken) phase corresponds to $\gamma>1$ ($\gamma <1$). 
Interestingly, this transition point $\gamma_{th}=1$ obtained from our numerical results is exactly  same as what we obtained by analyzing the $2\times 2$ matrix in the previous section. 

\begin{figure}
\includegraphics[width=0.46\textwidth]{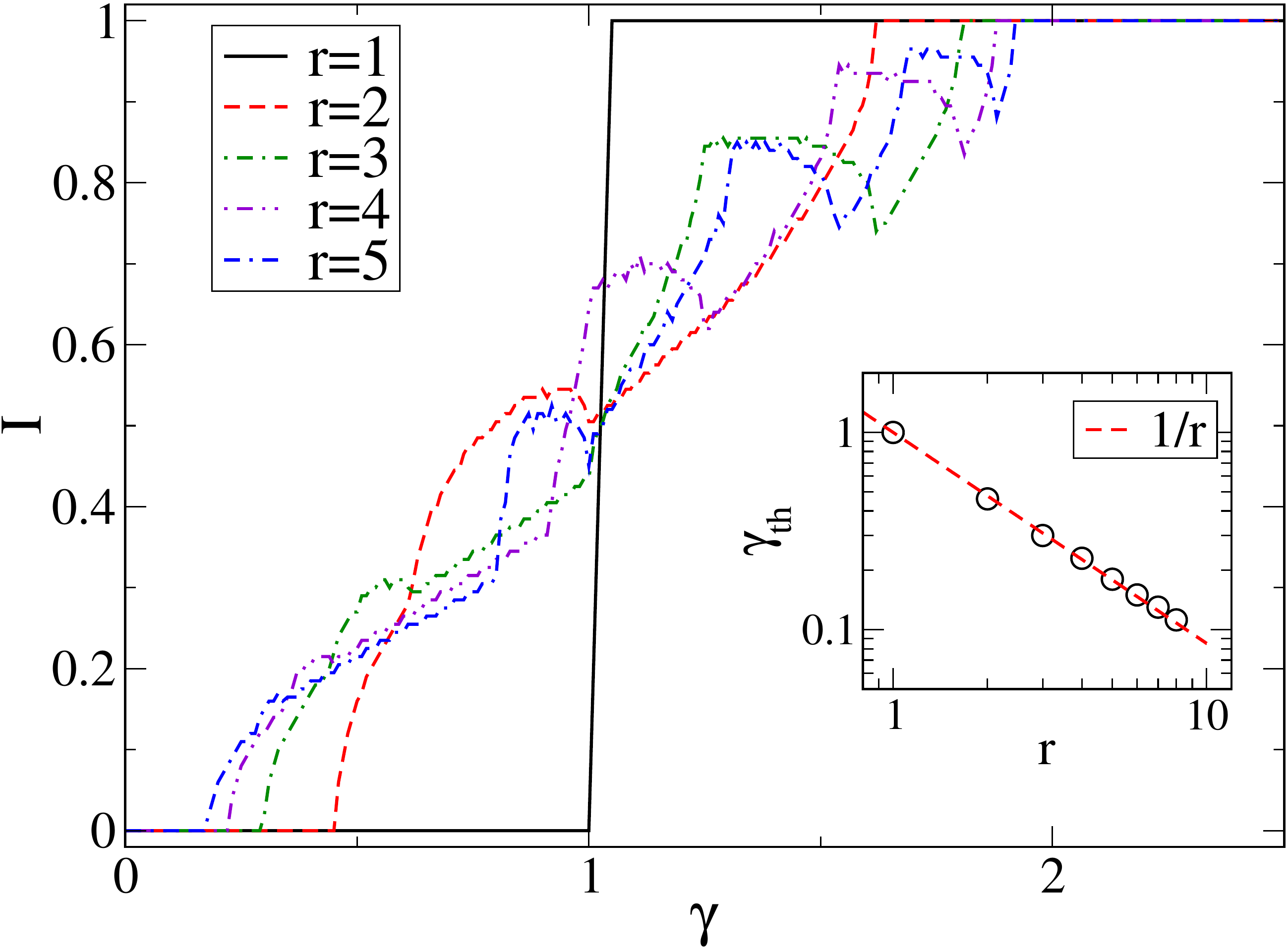}
\caption{Variation of the fraction of single-particle complex eigenvalues $I$
as a function of the hermiticity breaking parameter
$\gamma$ for $L=400$, Hamiltonian $H$
~\eqref{hamiltonian_pt4} for different values of $r$. Inset shows  $\gamma_{th}$ scales to zero 
with increasing $r$ as a power-law. Dashed line corresponds to $~1/r$.}
\label{fig1_extra}
\end{figure}

\textcolor{black}{ We also like to point out that $PT$ transition we have observed  for Hamiltonian~\eqref{hamiltonian_pt}
is unique in the sense that  here in the $PT$ broken phase all energy eigenvalues are complex, and hence at $\gamma=1$, the faction of complex eigenvalues $I$ jumps  from $O$ to $1$. However, this feature is just a manifestation of the fact that we have added the hermiticity breaking parameter only at sites $L/2$ and $L/2+1$. In the inset  
 of Fig.~\ref{fig1}, we have studied the Hamiltonian $\tilde{H}$ (see Eqn.~\eqref{hamiltonian_pt4}) where we have added the hermiticity breaking terms in 4 sites. We show that the $PT$ transition points $\gamma_{th}\simeq 0.45 < 1$ and also we find that here $I$ 
 does not jumps sharply from $0$ to $1$, in contrast there is a parameter regime where $I$ takes values between $[0,1]$. It implies that in that parameter regime,  a fraction of eigenvalues still remains real. However, as we increase the $\gamma$ all energy eigenvalues become complex. 
 \textcolor{black}{Interestingly, we also like to point out that for this model near $\gamma=1$, the variation of $I$ with $\gamma$ is non-monotonic. There is a parameter regime near $\gamma=1$ for which the number of complex eigenvalues decreases as we increase the hermiticity breaking parameter.} \textcolor{black}{Figure.~\ref{fig1_extra} shows the variation of $I$
 with $\gamma$ for the Hamiltonian Eqn.~\eqref{hamiltonian_pt4} for different values of $r$ and for $L=400$. We find indeed the $\gamma_{th}$ becomes much smaller as we increase $r$. Inset shows that the  $\gamma_{th}$ approaches to zero with $r$ 
 as a powerlaw. Interestingly we find as the number of Hermiticity-breaking terms increases, 
the region  where real and imaginary eigenvalues co-exist i.e. $0<I<1$,  also increase.} 
 Note that it is straightforward to show even analytically that the  $PT$ symmetric phase would not have been stable in the thermodynamic limit if we had added hermiticity breaking term at all sites [see Appendix.~\ref{appendixI} for more details].}

\begin{figure}
\includegraphics[width=0.46\textwidth]{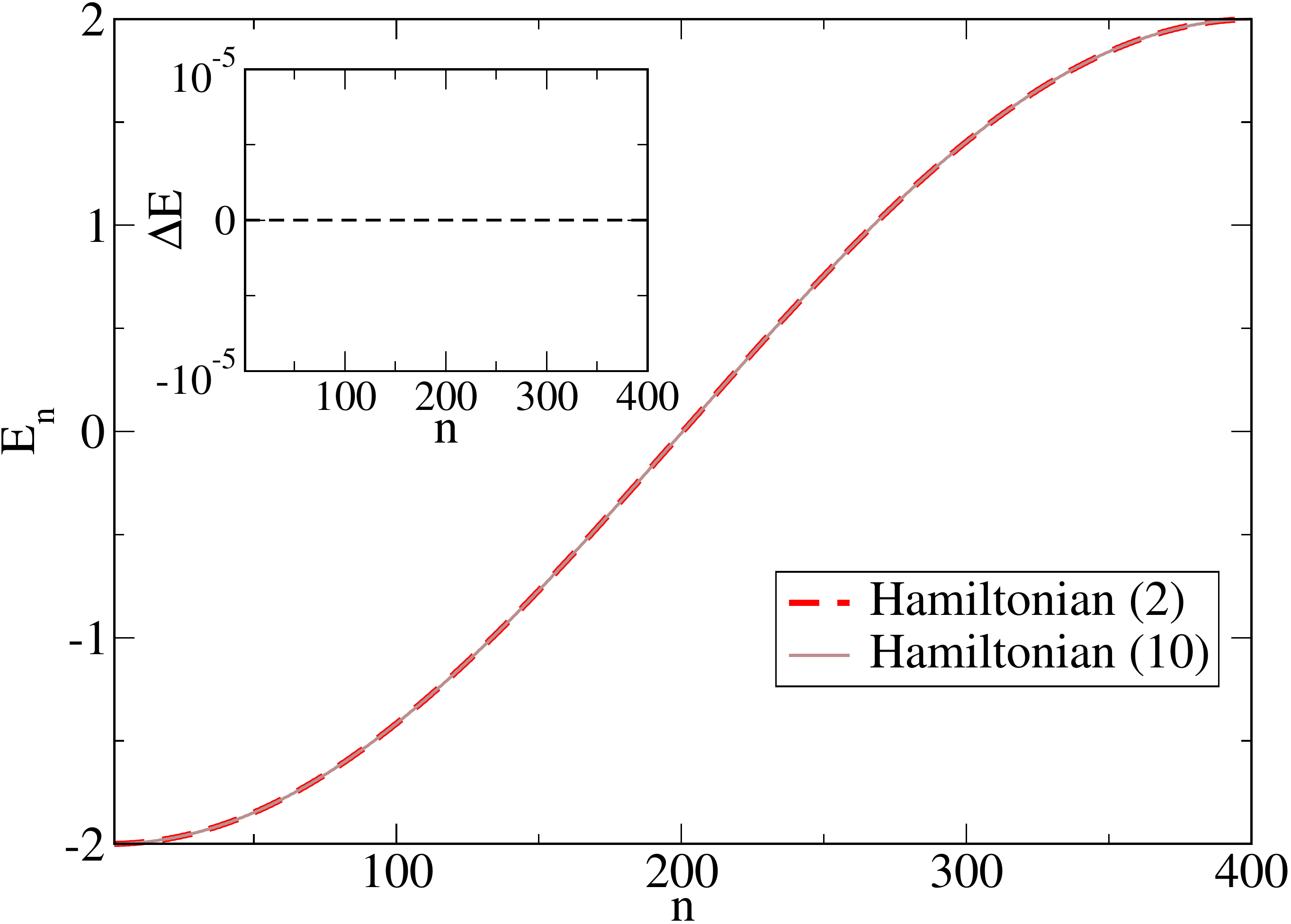}
\caption{Comparison of energy eigenvalues $E_n$ between Hamiltonian~\eqref{hamiltonian_pt} and Hamiltonian~\eqref{hamiltonian_hermitian1}  for 
$\gamma=0.5$ and $L=400$. Inset shows the differences of the energy eigenvalues between Hamiltonian~\eqref{hamiltonian_pt} and Hamiltonian~\eqref{hamiltonian_hermitian1}
are of the order of machine precision.  }
\label{energy1}
\end{figure}

\begin{figure}
\includegraphics[width=0.46\textwidth]{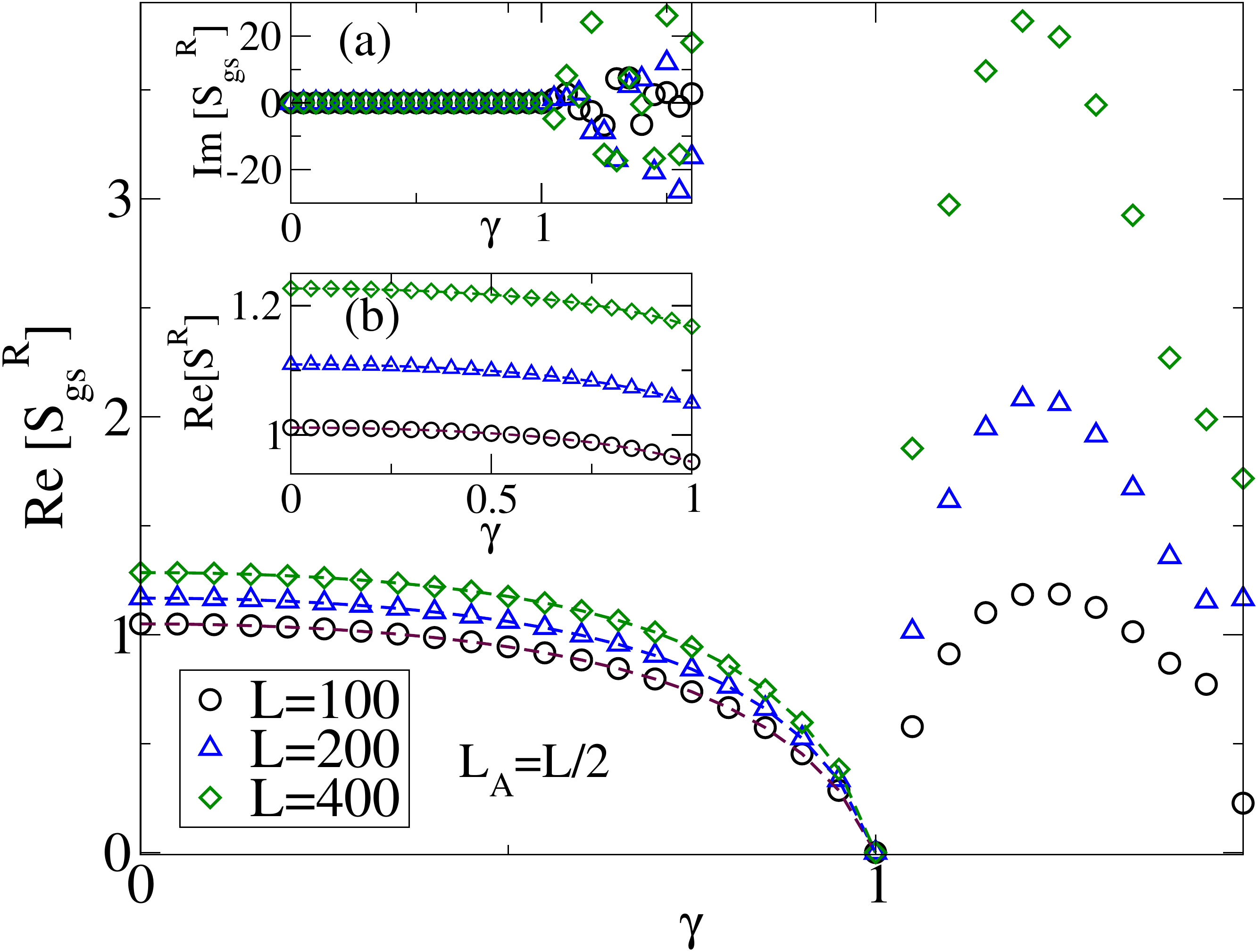}
\caption{  Variation of the real part of $S^{R}$ of the ground state as a function of $\gamma$ for $L=100$, 200, 400. 
We choose the subsystem size $L_A=L/2$. Solid dashed lines corresponds to ground state entanglement entropy of the Hermitian 
model $h$ ~\eqref{hamiltonian_hermitian1}. Inset (a) shows the variation of the imaginary part of $S^{R}$ of the ground state as a function of $\gamma$. Inset (b) shows the variation of the real part of $S^{R}$ of the ground state as a function of $\gamma$ for the subsystem size $L_A=L/4$. Solid dashed lines corresponds to ground state entanglement entropy of the Hermitian model $h$ ~\eqref{hamiltonian_hermitian1}. }
\label{fig2}
\end{figure}

\section{Entanglement entropy} \label{secV}
In this section we will discuss the many-body eigenstate entanglement entropy of the $PT$ invariant Hamiltonian \eqref{hamiltonian_pt}.
We note that a typical measure of the entanglement in a quantum system is bipartite von Neumann entanglement entropy
$S$ defined as, $S=-\Tr_{A}[\rho_A\ln\rho_{A}],$
where $ \rho_{A}=\Tr_{B}\rho $ is the 
reduced density matrix of a sub-system $A$ after dividing the system into two  adjacent  parts $A$  and $B$. 
$\rho$ is the total density matrix corresponding to the eigenstate  of the  system. 
For Hermitian system, $\rho=|E_n\rangle\langle E_n|$, where $E_n$s are many-body energy eigenstates. 
For non-Hermitian system, since left and right eigenvectors are not same, we have two choices to define 
the total many-body density matrix of the system, i.e. 1) $\rho_1=|R_n\rangle\langle R_n|$ and 2) $\rho_2=|R_n\rangle\langle L_n|$ and corresponding entanglement entropy are denoted by $S^{R}$ and $S^{LR}$ respectively ~\cite{gopalakrishnan2020entanglement,lee2020exceptional}. 
Note that since the Hamiltonian \eqref{hamiltonian_pt} is quadratic, the  ground state (also typical eigenstates)
entanglement entropy can be obtained from the one-body density matrix~\cite{herviou2019entanglement} in the similar spirit as one can do for Hermitian system \cite{lev1,lev2,lev3,lucas.2019,peschel2009reduced,PhysRevLett.125.180604}. 

Also, motivated by our analysis for two-sites model,  we conjecture that the  
Hermitian equivalent model $h$ corresponds to the non-Hermitian Hamiltonian $H$ is given by, 
\begin{eqnarray}
 h=-\sum _{j=1}^{L/2-1}(\hat{c}^{\dag}_j\hat{c}^{}_{j+1}+\text{H.c.})-\sum _{j=L/2+1}^{L-1}(\hat{c}^{\dag}_j\hat{c}^{}_{j+1}+\text{H.c.})\nonumber\\
 -(1-\gamma^{2})^{1/2}(\hat{c}^{\dag}_{L/2}\hat{c}^{}_{L/2+1}+\text{H.c.})\nonumber \\
 \label{hamiltonian_hermitian1}
\end{eqnarray}
Note that if we restrict ourself to $L=2$, the Hamiltonian $h$ becomes identical to $h^{2\times 2}$ (see Eqn.~\ref{h2times2}). \textcolor{black}{We also test our conjecture by comparing the single particle energy eigenvalues 
$E_n$ for the both Hamiltonians ~\eqref{hamiltonian_pt} and ~\eqref{hamiltonian_hermitian1} in the Fig.~\ref{energy1}. We find an excellent agreement between them as shown in Fig.~\ref{energy1}.}

Next, our goal is to analyze the entanglement entropy of the ground state and as well as the typical excited states for both the models $H$ (see Eqn.~\ref{hamiltonian_pt}) and $h$ (see Eqn.~\ref{hamiltonian_hermitian1}),  and compare their results.

\subsection{Ground state}
We first focus on the ground state entanglement entropy of the non-Hermitian Hamiltonian $H$. 
Usually the many body ground state of a non-interacting system is obtained by populating the lowest energy  single particle states one by one. The similar method can also 
be used for the non-Hermitian system in the $PT$ symmetric phase given that the energy eigenvalues 
are purely real. In case of $PT$ broken phase, given that the eigenvalues can be complex, there is a bit of ambiguity regarding in which order we should populate the single particle states to get the desired many-body ground state. However, we arrange the eigenvalues  sorting by its real parts and obtain the many-body ground state by populating one by one the single-particle states whose real parts of the eigenvalues are the lowest. \textcolor{black}{Given that we have always 
restricted ourself to the even number of particles in the system, this particular choice of the many-body ground state ensures that many-body ground state energy always remains real even in the $PT$ broken phase. Moreover, this working definition of the many-body ground state can also be ``analytically continued'' in the $PT$ symmetric 
phase.}

Figure.~\ref{fig2}
shows the variation of the ground state entanglement entropy obtained from the right eigenvectors. In the main panel of 
Fig.~\ref{fig2}, we plot the real part of $S^{R}$ as a function of $\gamma$. Interestingly, we find that the $S^{R}$ decreases
monotonically as a function of $\gamma$ when $\gamma < 1$, i.e. $PT$ symmetric phase. Remarkably, this value goes to zero 
at the transition point, i.e. $\gamma=1$. 
The solid dashed lines in the main panel of Fig.~\ref{fig2} corresponds to the ground state entanglement entropy of the Hermitian
Hamiltonian $h$. It shows an excellent agreement with the real part of  $S^{R}$ in the $PT$ symmetric phase. 
In the inset (a) of Fig.~\ref{fig2}, we also plot the variation of the imaginary part of the $S^{R}$, we find that while it is zero in the $PT$ symmetric phase, for $\gamma >1$ (in the  broken $PT$ phase)  the imaginary part of $S^{R}$ can have non-zero values.  \textcolor{black}{We also like to emphasize that Re[$S^{R}$] vanishes at $\gamma=1$  is due to the fact that we choose $L_A=L/2$. That becomes even more apparent from our Hermitian equivalent Hamiltonian ~\eqref{hamiltonian_hermitian1}. At $\gamma=1$, Hamiltonian  ~\eqref{hamiltonian_hermitian1}
is just two  separate systems of length $L/2$ who don't talk to each other. Hence, the entanglement entropy is bound to be zero if we make a cut in the middile. 
However, if we choose  $L_A=L/4$, at $\gamma=1$, the entanglement will not be zero. In the inset (b) of Fig.~\ref{fig2},
it has been expliclitly shown. However, the variation of the entanglement entropy  in the $PT$ symmetric phase can still
be captured by the Hermitian equivalent Hamiltonian Eqn.~\ref{hamiltonian_hermitian1}.}

\begin{figure}
\includegraphics[width=0.46\textwidth]{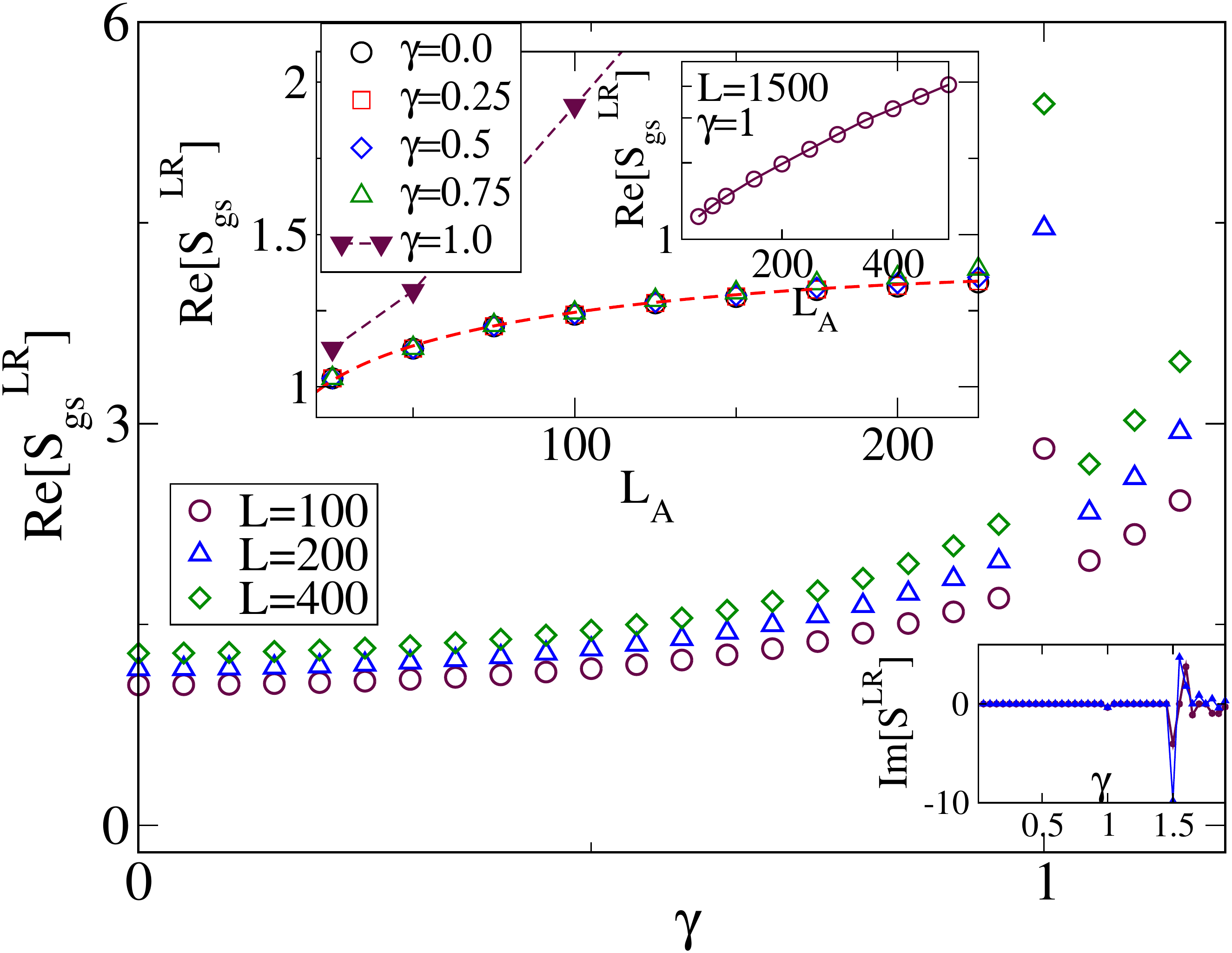}
\caption{
In the main panel we show the variation of the real part of $S^{LR}$ of the ground state as a function of $\gamma$ for $L=100$, 200, 400. We choose the subsystem size $L_A=L/2$.  Inset shows the variation of the real part of $S^{LR}$ with $L_A$ for $\gamma=0$, 0.25, 0.5, 0.75, 1. We keep $L=600$ fixed. Red dashed line is the best fit where the fitting function is $\frac{1}{6}\ln[\sin(\pi L_A/L)]$+ const. Another inset shows the variation Re[$S^{LR}$] with $L_A$ ( for fixed $L=1500$) for $\gamma=1$ in the semi-log scale. Inset in the right-bottom corner shows the variation of imaginary part of $S^{LR}$ with $\gamma$.    }
\label{fig3}
\end{figure}

\begin{figure}
\includegraphics[width=0.46\textwidth]{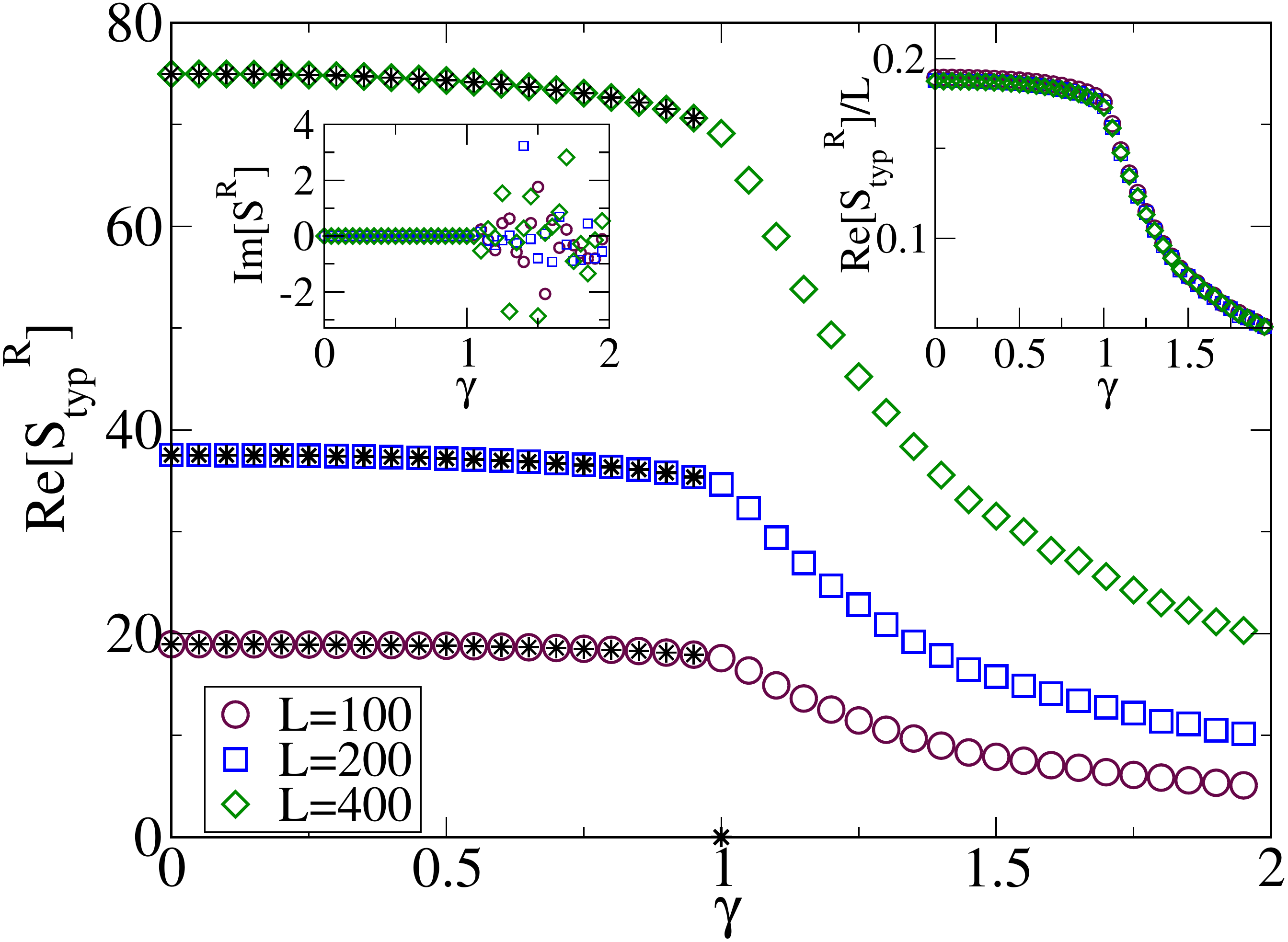}
\caption{ Variation of Re$[S^{R}]$ of typical eigenstates of the Hamiltonian~\eqref{hamiltonian_pt} as a function of $\gamma$ for $L=100$, 200, 400.
for fixed $L_A=L/2$. Black symbols correspond to the entanglement entropy of  a typical state for the Hermitian 
Hamiltonian $h$. 
Top right Inset  shows  a nice data collapse when we re-scale Re$[S^{R}]$ by Re$[S^{R}]/L_A$. Another inset shows the variation of imaginary part of $S^{R}$ with $\gamma$.  }
\label{fig4}
\end{figure}

\begin{figure}
\includegraphics[width=0.46\textwidth]{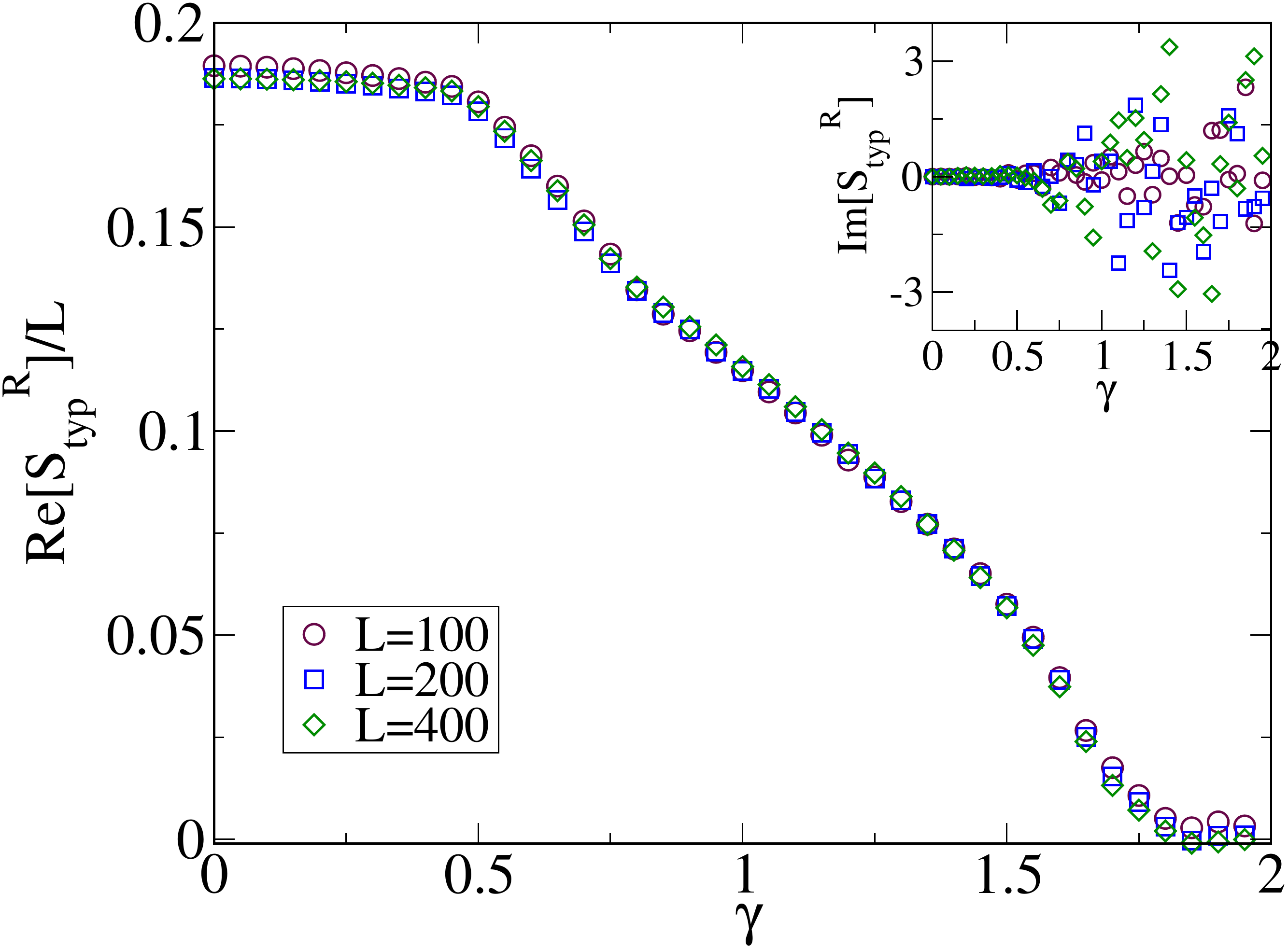}
\caption{ Variation of Re$[S^{R}]/L$ of typical eigenstates of the Hamiltonian~\eqref{hamiltonian_pt4} as a function of $\gamma$ for $L=100$, 200, 400.
for fixed $L_A=L/2$, which shows nice data collapse. 
Inset shows the variation of  Im$[S^{R}]$ with $\gamma$ for the same model. }
\label{fig4_extra}
\end{figure}

\begin{figure}
\includegraphics[width=0.46\textwidth]{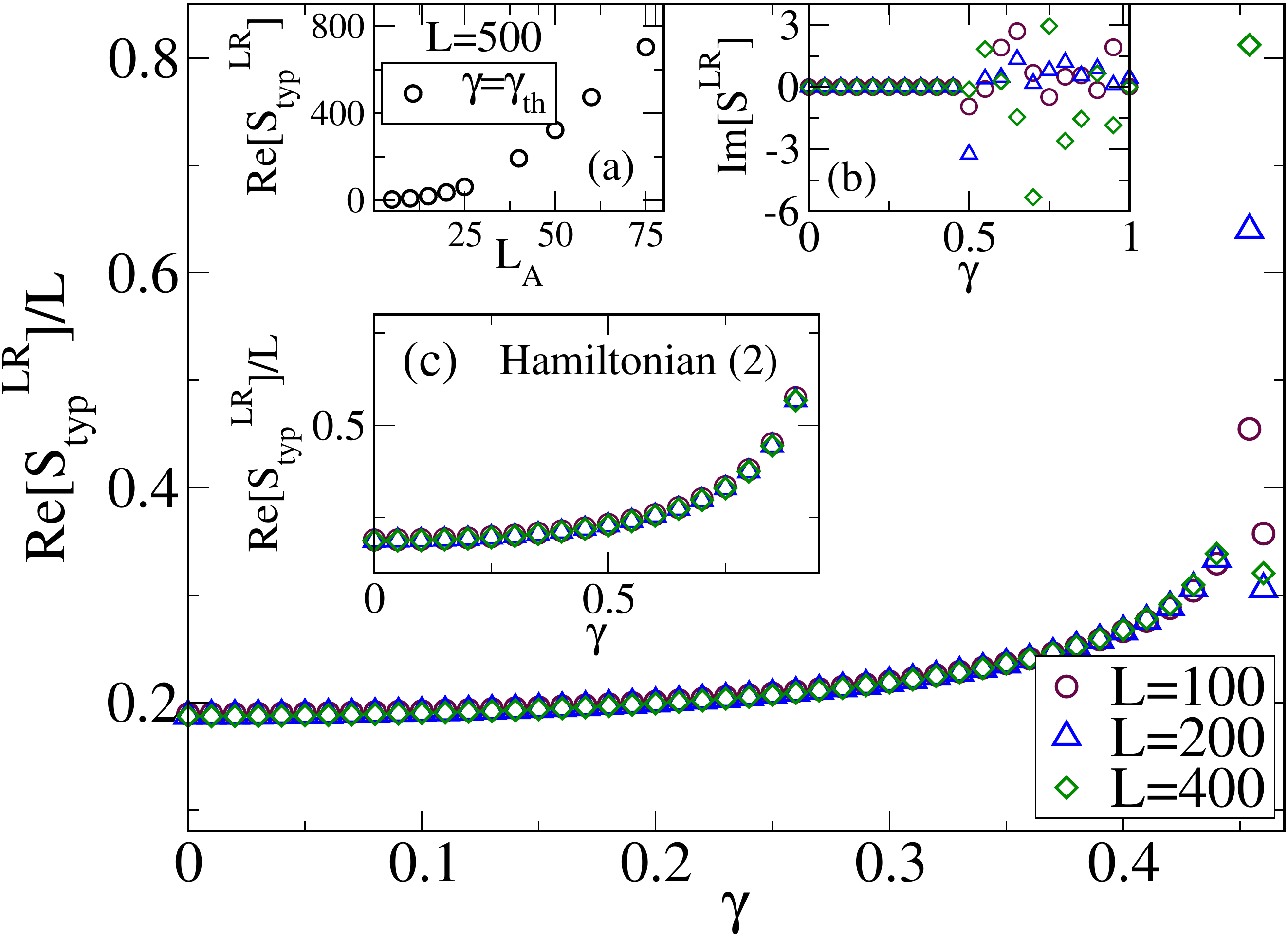}
\caption{Variation of Re$[S^{LR}]/L$ of typical eigenstates as a function of $\gamma$ for $L=100$, 200, 400
for fixed $L_A=L/2$ and
for the Hamiltonian $\tilde{H}$ ~\eqref{hamiltonian_pt4}.
Data collapse is observed for $\gamma < \gamma _{th}\simeq 0.45$. 
Insets (a) shows the variation of  Re$[S_{typ}^{LR}]$ with $L_A$ for $L=500$ and $\gamma$ = $\gamma _{th}$. 
Inset (b) shows the variation of the imaginary part of $S^{LR}$ with $\gamma$. Inset (c) shows the data collapse of $S^{LR}/L$
for Hamiltonian Eqn.~\ref{hamiltonian_pt} in the $PT$ symmetric phase i.e. $\gamma<1$.
}
\label{fig5}
\end{figure}


Next we investigate how $S^{LR}$  behaves as a function of Hermiticity breaking parameter $\gamma$. We again focus on the ground state. In the main panel of Fig.~\ref{fig3}, we plot the real part of $S^{LR}$ with $\gamma$ for different values of $L$, we keep the subsystem size  $L_A=L/2$  fixed. We find that in the $PT$ symmetric phase and for a fixed value of $\gamma$,  the real part of $S^{LR}$ increases with system size. 
First we focus on $PT$ symmetric phase, i.e. $\gamma<1$. For $\gamma=0$ we know that the Hamiltonian $H$ is gapless, which implies that the ground state can be described by a CFT of central charge $c$ . Hence, we expect that the entanglement entropy  $S$ should scale as $S=\frac{c}{6}\ln[\sin(\pi L_A/L)]+$const ~\cite{chang2020entanglement}. It is well known that  the central charge corresponding to the Hamiltonian $H_0$ is $c=1$. In the inset of Fig.~\ref{fig3}, we show that indeed for $\gamma=0$, $S$ obeys expected logarithmic scaling. Strikingly we find that the same scaling exists for $S^{LR}$ in the $PT$ symmetric phase as well. Next, we investigate the behavior of $S^{LR}$ at the phase transition point i.e. $\gamma=1$. We observe that at $\gamma=1$   the value of Re[$S^{LR}$] is much larger compared to its value in the $PT$ symmetric phase.  In the inset we show the variation of Re[$S^{LR}$] with $L_A$ for a fixed value of $L$ and we find that $S^{LR}$ actually  diverges exponentially with $L_A$  at the phase transition point. \textcolor{black}{We also show the variation of the imaginary part of $S^{LR}$ in  the right-bottom corner inset of Fig.~\ref{fig3} and find that in the $PT$
symmetric phase imaginary part of $S^{LR}$ is zero, whereas in the $PT$ broken phase it can have non-zero value.}

\subsection{Typical eigenstates}
After investigating the ground state entanglement entropy, now we study the entanglement entropy of the typical states.
Again we investigate separately both $S^{R}$ and $S^{LR}$. In each configuration, we randomly populate $L/2$ numbers of 
particles in total $L$ number of single particle states. Then we average over $1000$ different configurations ~\cite{lev1,lev2,lev3,modak.prr}. 
Figure.~\ref{fig4}, shows the variation of the real part of $S^{R}$ of a typical state with $\gamma$ for the Hamiltonian $H$ 
\eqref{hamiltonian_pt}. We see that in the $PT$ unbroken
phase, the magnitude of $S^{R}$ does not change significantly, it remains almost the same as the value obtained for $\gamma=0$. 
However, for $\gamma>1$, the value monotonically decreases. 
Also, similar to the ground state, Re[$S^{R}$] in the $PT$ symmetric phase is almost the same as the entanglement 
obtained for the typical eigenstates of Hermitian Hamiltonian $h$ ~\eqref{hamiltonian_hermitian1}. 
In the inset, we show the data collapse when we re-scale Re[$S^{R}$] by Re[$S^{R}$]$/L_A$, which indicates the signature of the volume law.\textcolor{black}{We strengthen our claims by investigating the Hamiltonian Eqn.~\eqref{hamiltonian_pt4} in Fig.~\ref{fig4_extra} as well.}

\textcolor{black}{Next, we investigate the real part of $S^{LR}/L$ in Fig.~\ref{fig5} for the Hamiltonian 
$\tilde{H}$ ~\eqref{hamiltonian_pt4}. We again find a nice data collapse in the $PT$ symmetric phase i.e. $\gamma<0.45$, which indicates
the volume entanglement. However, at the transition point $\gamma\simeq 0.45$, the volume law scaling breaks down. At the point of 
phase transition even after we rescale Re$[S^{LR}]$ by Re$[S^{LR}]/L$, the value increases with $L$, that suggests the scaling 
is much faster than the volume law.
In the inset (a) of Fig.~\ref{fig5}, we plot the variation  Re$[S^{LR}]$ for the Hamiltonian $\tilde{H}$ at the $PT$ transition point  with $L_A$ for a fixed $L=500$, and  we confirm
that the scaling of Re$[S^{LR}]$ with system size is much faster than the volume law scaling observed in the $PT$ symmetric phase.}\textcolor{black}{We also show the imaginary part of $S^{LR}$ for typical state in the inset (b), which shows 
that similar to our previous findings in the $PT$ symmetric (broken) phase Im[$S^{LR}$] is zero (non-zero).
Note that the  similar behavior has been observed even for the Hamiltonian $H$ ~\eqref{hamiltonian_pt}. In the inset (c) of Fig.~\ref{fig5}, we show the nice data collapse for $S^{LR}/L$ in $PT$ symmetric phase i.e. $\gamma<1$.}

\section{Conclusions}\label{secVI}
In this paper, we investigate the entanglement properties of the ground state and a typical excited state of a non-interacting non-Hermitian lattice model which is invariant under $PT$ transformation. The model we had looked into, has two parts. The Hermitian part is described by a fermionic system having just nearest-neighbor hopping, while we add the Hermiticity breaking terms (also known as gain-loss terms) only at the two sites (four sites for the Hamiltonian $\tilde{H}$) which are situated in the middle of the lattice. 
\textcolor{black}{Usually in Hermitian systems if one adds some local pertubations, it does not alter significantly  the extensive quantities like energy of the system,  strikingly here we find that even though we have added the hermicity breaking terms only in a finite number of sites (in 2 sites for Hamiltonian ~\eqref{hamiltonian_pt} and in 4 sites for Hamiltonian ~\eqref{hamiltonian_pt4}), but it significantly modifies the properties of the 
system.}
Most importantly, we find that the non-Hermitian model shows $PT$ phase transition as we change the Hermiticity breaking 
parameter. 

Next, we analyze the entanglement properties of different phases of this model. We find that in the $PT$ symmetric phase, the entanglement entropy obtained from only the right eigenvetors, are same as the entanglement entropy obtained for the equivalent 
Hermitian model. We find this result extremely non-trivial, given that the equivalence between the non-Hermitian and Hermitian model implies they have same set of energy eigenvalues, but it does not imply that the eigenvectors of these two models are the same. Hence, 
It is not at all obvious that entanglement entropy of the eigenstates of these two models should be the same. 

Another interesting finding of our work is that the entanglement entropy obtained by considering both left and right vectors, 
diverges exponentially with system size at the $PT$ phase transition point. On the other hand, in the $PT$ symmetric phase, the ground state entanglement entropy scales logarithmically with subsystem size, which is an  evidence that presumably  low energy states of this 
$PT$ symmetric non-Hermitian system, might also be described by CFT. We also find that the typical excited states are volume law entangled. 
\textcolor{black}{We would also like to point out that we have repeated all our calculations for the models 
where we have added the hermiticity breaking parameters even at more than $4$ sites. We find that even though the the 
results changes quantitatively, qualitative features of the  entanglement entropy remains un-altered. Interestingly 
the $PT$ phase transition point $\gamma_{th}$ approach to zero  as we increase the number of non-Hermitian sites, hence 
we only show the results for  the Hamiltonian $H$ ~\eqref{hamiltonian_pt} and $\tilde{H}$ ~\eqref{hamiltonian_pt4}.}

Our future plan is to understand the effect of interaction and disorder in such systems and study 
how they modify the $PT$ phase transitions. 
Recently, there have been efforts to investigate non-Hermitian Many-body localized phase ~\cite{PhysRevLett.123.090603}, it will be interesting to 
investigate the $PT$ symmetric system in the same shade of lights. 

\section{Acknowledgements} 

RM acknowledges the support of DST-Inspire fellowship, by the
Department of Science and Technology, Government of India. The authors would like to thank Debraj Rakshit for many stimulating discussions.

\appendix
\section{}\label{appendixI}

Here we investigate a similar model as ~\eqref{hamiltonian_pt}, but we add complex on-site potential in each site ~\cite{PhysRevA.84.024103}.
The model is described by the following Hamiltonian, 
\begin{eqnarray}
\tilde{H}=H_0+i\gamma \sum_{j}(-)^j\hat{n}_j \nonumber \\
\label{hamiltonian_pt_extensive}
\end{eqnarray}
It is straightforward to verify that the Hamiltonian \eqref{hamiltonian_pt_extensive} also invariant under 
$PT$ transformation, i.e. $[\tilde{H},PT]=0$ 

Thus, we arrive at the stationary discrete Schrodinger equation,
\begin{eqnarray}
 E\psi_j=\psi_{j+1}+\psi_{j-1}+ i\gamma (-1)^j \psi_j\nonumber \\
\label{schorindger eq}
\end{eqnarray}
We assume a trial solution $\psi_j=Ae^{ijk}+Be^{-ijk}$. Using the boundary condition 
$\psi_0=\psi_{L+1}=0$, one obtains $A=-B$ and $k=s\pi/(L+1)$, with $s=0,1$, $,\cdots$ $L-1$. 
Inserting the trial solution into the Eqn.~\eqref{schorindger eq}, it is straightforward to obtain 
the energy eigenvalues, which is given by
\begin{eqnarray}
E^{2}=4\cos^{2}k-\gamma ^{2}.
\label{energy}
\end{eqnarray}
From Eq.~\eqref{energy}, one gets that all eigenvalues are real for $\gamma < 2\cos k$ for any
value of $s$. Since the smallest value of $\cos k$ occurs for $s = L/2$,
the condition of a completely real spectrum is
\begin{eqnarray}
\gamma < \gamma _{th}=2\cos\bigg[\frac{L\pi}{2(L+1)}\bigg]\simeq \pi/L
\label{threshold}
\end{eqnarray}
Hence, in the thermodynamic limit the Hamiltonian~\eqref{hamiltonian_pt_extensive} does not have a true $PT$ 
unbroken phase.

\section{} \label{appendixII}

In this section we restrict the  Hamiltonian $\tilde{H}$ Eqn.~\eqref{hamiltonian_pt4}   to only a lattice of  four sites.  
We show the details of the numerical calculation, which one can use to obtain the Hermitian equivalent Hamiltonian for this system. The outline of the calculation is already presented in the main text in the section sec.~\ref{secIII}. 
The Hamiltonian  $4\times 4$ is represented in the matrix form as 
\begin{eqnarray}
H^{4\times 4}=
 \begin{pmatrix}
 i\gamma & -1 & 0 & 0 \\
 -1 & -i\gamma &-1& 0 \\
  0&-1 & i\gamma & -1 \\
  0&0&-1 & -i\gamma  \\
\end{pmatrix}
\label{h4cross4}
\end{eqnarray}
We choose $\gamma=0.2$, for which all energy eigenvalues are completely real i.e. $E_1=-1.60563$, $E_2=1.60563$, $E_3=0.584779$,and  $E_4=-0.584779$. 
Hence, it belongs to the $PT$ symmetric phase. Simultaneous eigenvectors of $H^{4\times 4}$ and $PT$ are 
$|E_1\rangle=(0.373182,0.59919+0.07463 i,0.603821,0.37032+0.0461278 i)^T$, $|E_2\rangle=(0.373182,-0.59919+0.07463 i,0.603821,-0.37032+0.0461278 i)^T$, $|E_3\rangle=(0.585094+0.200108 i,-0.382172,-0.361608-0.123674 i,0.618368)^T$, and $|E_4\rangle=(0.618368,0.361608+0.123674 i,-0.382172,-0.585094-0.200108 i)^T$, where $PT|E_{1,3}\rangle=|E_{1,3}\rangle$, and $PT|E_{2,4}\rangle=-|E_{2,4}\rangle$. 
Now, it is straightforward to obtain $C$ operator, where $C^{2}=I$, and $[C,H^{4\times 4}]=0$, it reads as, 
\begin{eqnarray}
C=
 \begin{pmatrix}
 -0.2819 i& 0.92331 & 0.097245 i & -0.486228 \\
 0.92331 & 0.184663 i &0.437088& -0.097245 i \\
  0.097245 i &0.437088 & -0.184663 i & 0.92331  \\
  -0.486228&-0.097245 i& 0.923316 & 0.2819 i  \nonumber
\end{pmatrix}
\label{cmat}
\end{eqnarray}
Given that $C=e^QP$ and the parity operator reads as, 
\begin{eqnarray}
P=
 \begin{pmatrix}
 0 & 0 & 0 & 1 \\
 0 & 0 &1& 0 \\
  0&1 & 0 & 0 \\
  1&0&0 & 0 \\

\end{pmatrix}
\label{p4}
\end{eqnarray}
it is trivial to obtain $h^{4\times 4}=e^{-Q/2}H^{4\times 4}e^{Q/2}$, which is represented as, 
\begin{eqnarray}
h^{4\times 4}=
 \begin{pmatrix}
 0 & -0.979579 & 0 & -0.0206343\\
 -0.979579 & 0 &-1.00021& 0 \\
  0&-1.00021 & 0 & -0.979579 \\
  -0.0206343&0&-0.979579& 0  \nonumber 
\end{pmatrix}
\label{hmat}
\end{eqnarray}
where, $h^{4\times 4}$ is a Hermitian matrix and it is straight forward to check that it's eigenvalues are same as the eigenvalues of 
$H^{4\times 4}$, i.e. $E_1$, $E_2$, $E_3$,and  $E_4$.  Note that  one can perform the similar calculation for any $PT$  symmetric non-Hermitian system, and obtain the the Hermitian equivalent Hamiltonian.

\bibliography{reference}

\end{document}